\documentclass[aps,prb,preprint,groupedaddress]{revtex4}
\usepackage{graphicx}

\begin{document}
\renewcommand\floatpagefraction{.001}
\makeatletter
\setlength\@fpsep{\textheight}
\makeatother
\setcounter{totalnumber}{1}

\title{ Sum of squares decompositions of  potential energy surfaces }

\author{Martin Burke and Sophia N. Yaliraki}
\email[]{s.yaliraki@imperial.ac.uk}
\affiliation{Department of Chemistry, Imperial College London, 
 South Kensington Campus,
  London, SW7 2AZ, United Kingdom}

\date{\today}

\begin{abstract}
The difficulty in exploring potential energy surfaces, which are nonconvex, stems from the presence of many local minima, typically separated by  high barriers and  often disconnected in configurational space. We  obtain the global minimum on model potential energy
surfaces without sampling any  minima a priori. Instead, a different  problem is derived, which is convex and hence easy to solve, but  which is guaranteed to either  have the same solution or to  be a lower bound to the true solution. A  systematic way  for improving the latter solutions is also given. Because many nonconvex problems are projections  of higher dimensional convex problems, Parrilo\cite{parrilo_thesis00,parrilo_book03} has recently showed that by obtaining   a  sum of squares decomposition of the original problem, which can be  subsequently transformed to 
a semidefinite programme a large class of non-convex problems can be solved efficiently. The 
semidefinite duality formulation also  provides a  proof that the global minimum of 
the energy surface has either been found exactly or has  been bounded from below.  It additionally   provides physical insight into the problem through a geometric interpretation. The sum of squares polynomial representation of the potential energy surface  may further reveal   information 
about the nature of the potential energy surface.    
We demonstrate the applicability of this 
approach to low dimensional potential energy landscapes and discuss its 
merits and shortcomings.
\end{abstract}

\maketitle


\section{Introduction}
The thermodynamic and kinetic  properties of atomic and molecular assemblies in physics, chemistry and biology are 
often linked to their underlying potential energy surface (PES)\cite{berry93,wales99,tharrington02}. Even in the simplest cases, exploring potential energy surfaces or even 
just important regions becomes  computationally demanding very rapidly. 
This is primarily due to the non-convexity of the   underlying potential energy surface. A key step in understanding thermodynamic  properties of  systems as diverse as proteins, crystals or nanoscale clusters is the identification 
of the global minimum. Definite proof typically requires  exhaustive sampling of every other minimum in configurational
space, a prohibitive task given that the number of 
local minima on the surface grows exponentially with the number of particles in 
the system~\cite{berry93,jordan93}. 

A series of  approaches have been proposed to tackle this problem with various degrees of 
success. Most stochastic  methods are variations
 of the  Monte Carlo (MC) method with additional steps  introduced
 to aid barrier crossing between local minima
 \cite{kirkpatrick83,frantz90,grubmueller95}, although
 genetic-based algorithms\cite{rabow96,deaven96} have also been used.
   Deterministic  approaches  usually derive from global optimisation 
techniques~\cite{floudas92,ratner04,straub98} or molecular dyanamics~\cite{wales_berry95}. Methods that deform the potential in order
   to eliminate local minima include  hypersurface deformation
   techniques \cite{piela89}, such as the distance 
scaling method \cite{pillardy95}, 
 and the  very successful MC basin-hopping method~\cite{wales97}.

A commonality in most  methods is that they  require
extensive sampling of the local minima of the potential energy surface itself prior to
finding the global minimum. 
Not only is this process time consuming, sensitive to initial conditions, and susceptible to trapping due to high barriers,  
it crucially provides no proof that a given
point on the PES is indeed the global minimum other than that it is the lowest 
visited to date by the algorithm. It is typically not possible to
evaluate how close such a minimum would be to  the correct answer to
further guide the search. Furthermore,  all  methods rely to a greater
or lesser extent on the topology of the PES being 'well 
behaved' in the sense that the global minimum is  connected  
to other low lying local minima so that it can be reached from them. 

Here we outline a different strategy  that  addresses 
the problem of locating the global minimum without sampling other minima in 
configurational space. As a result, it is  independent of initial conditions and of any  requirement that the minimum  be closely related to any other. 
Either the exact result is found together with a proof or a lower bound is obtained. In no case is the original  function  deformed.  A key idea is that instead of sampling the underlying non-convex function, we approach it from below. A different problem is sought, whose {\it solution} coincides with the solution of the problem we are interested in.  For this approach to be successful, it must be coupled both with a rigorous and systematic way to find such problems that can be easily solved, and, with a way to evaluate how close the two solutions are. An important mathematical concept that provides such a link  is that many nonconvex problems turn out to be projections of higher dimensional convex problems. The great advantage of the latter is that they are  efficiently solved since the local minimum is the global minimum. Furthermore, they also allow for physical insight into the  system because they possess duality properties. This duality provides geometric interpretations to the original problem and allows us to monitor how close the solution of the derived problem is to the true solution we are seeking. This is coupled with a systematic way to improve on our non-exact solutions. The combination of these features  makes this approach appealing because  physical insight is revealed about the system while seeking  the solution itself.  The approach, originally developed by Parrilo~\cite{parrilo_thesis00} for semialgebraic geometry problems in robust control,  is applicable to polynomial systems and is based on obtaining a  
sum of squares (SOS) 
representation of the original function,  which turns out to be convex in the coefficient space of the function, and, hence efficiently computed through semidefinite programming techniques. 
In Sec~\ref{Theory} we give the main concepts of the theory and its relation to physical observables and  in Sec~\ref{Results} we apply it to a series of low dimensional examples that exemplify a range of  difficulties encountered in  complex landscape problems  and which  highlight the  different capabilities and versatility of this method.  In Sec~\ref{Discussion} we discuss the advantages and shortcomings of the approach.   

\section{Theory}\label{Theory} 
\subsection{SOS decompositions}
A different approach to locating the global minimum $U^*$ of a
multivariable polynomial, $U(x_i,...x_n)\equiv U$ of degree $d$ is to look for  the largest real number $\lambda$ such that 
$U-\lambda$ can be decomposed as a SOS\cite{shor87,parrilo_thesis00,parrilo_book03}, that is 
\begin{equation}
U-\lambda=\sum^r_{i=1} p_i^2 \geq 0
\label{shor_eq}
\end{equation}
 where $p_i$ are polynomial functions of degree $d/2$ in the variables of $U$ (Fig. 1, steps 1-3).
Because being a SOS is a sufficient condition for a function to be nonnegative
(although not a necessary one), it is clear that if a SOS decomposition exists,  $\lambda$  will always be a lower bound to the original function's global minimum,  
$U^*-\lambda^*\geq f^{SOS}$ where  $^*$ denotes the
optimal and  $f^{SOS}$ the SOS  decomposition~\cite{shor87}. Indeed, both $\lambda$ and $f^{SOS}$ are guaranteed to bound the original problem from below~\cite{shor87}.  The 
next step then becomes to obtain the SOS decomposition together with $\lambda$ that gives the closest answer to our original problem.

 A central idea is the systematic lifting of a polynomial function in its {\it coefficient}  space (rather than its variables) to a SOS function. An important result from real algebraic geometry, the Positivstellensatz\cite{stengle74}, provides such a lifting.

The problem becomes tractable  by additionally exploiting the theorem~\cite{parrilo_thesis00} that for globally non-negative polynomial functions, there exists a positive semidefinite matrix 
$Q$ such that:
\begin{equation}
U(x)-\lambda=f^{SOS}=z(x)^TQz(x), \nonumber \qquad \textrm{where }Q\succeq 0,
\label{eq:gram_matrix}
\end{equation}
so that  the  original problem of Eq.~\ref{shor_eq} can now be formulated algebraically.   $z(x)$ is the vector of monomials of $U$ with degree less than or
equal to half the degree of U  and provides a basis for the SOS
function decomposition.  Since the variables in $z$ are not algebraically independent, the matrix Q is not unique and may be positive semidefinite in some representations but not in others. The problem now becomes a  search for a  symmetric 
positive definite matrix Q  that satisfies the constraints imposed by Eq.~\ref{eq:gram_matrix}.    Note  that the original non-convex  
problem in the variables of U has been mapped to a different problem, namely obtaining the polynomial coefficients of the SOS representation, which is 
equivalent to obtaining the matrix elements of  Q. This latter problem, although  in  higher dimensions,  is 
now convex and amenable to  efficient solutions  by SDP. Additionally, 
the minimisation of the potential is   carried out concurrently with the 
  search for the SOS coefficients.  The theoretical elegance of this methodology coupled with efficient computational techniques, allows for   
SOS decompositions to be found in polynomial time (As long as either the degree of the polynomial or the number of variables are fixed) and additionally  provides the 
proof that the result is indeed a lower bound to the global minimum.  If the answer is not exact,  the gap from the true solution is known. 

\subsection{SDP and  Duality}

SDPs\cite{vandenberghe96} are a class of optimisation problems over positive definite matrices solvable by  polynomial time 
algorithms such as the very
successful primal-dual interior point methods \cite{nesterov88}. Such methods  compute the optimal solution as well as provide 
certificates (proofs) of optimality. These certificates are based upon a  duality 
theory in which two problems are solved simultaneously. The dual problem solution provides a lower bound on the primal 
problem.  The difference between the primal and dual solutions, referred to as the duality gap, proves whether an exact solution has been found or provides the lower bound to the solution, which is away by  at most the magnitude of the duality gap.  

 The original polynomial minimisation can be recast as the SDP  given in  Eq. 4 in  Fig. 1. In our case, the primal problem gives  the lower bound to the global 
minimum as well as the SOS decomposition, while the dual provides the location of this bound in the configurational 
space. In the mathematical community the focus is typically on the 
primal, with the dual  only used to provide the duality gap. However, as we show here  there is relevant and  interesting 
physics in the dual problem as well. 

\subsection{Convexity and duality}

\begin{figure}[h!]
\includegraphics[width=12cm]{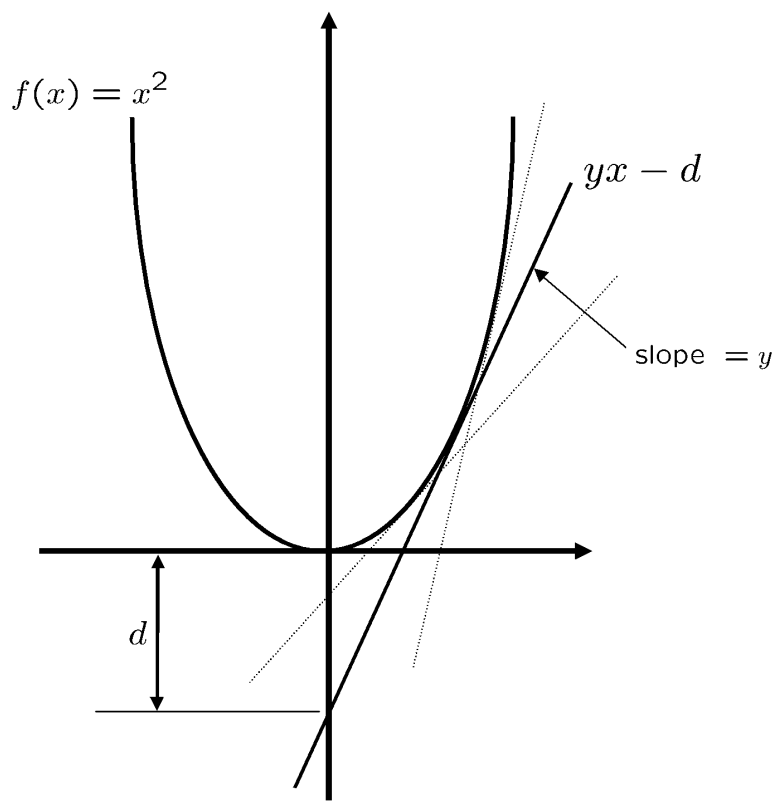}
\caption[Convex duality in Legendre transforms]{How convexity and
 duality are related. A set of lines with slope y lie below the
 parabola $f(x)=x^2$, as long as $yx-d\leq x^2$ where $d$ is the depth. The line will touch tangentially  the parabola when $d=1/4y^2$. Each point on the parabola can be represented as  a line which touches the parabola with different slopes y.  The envelope of these lines  reproduces the parabola itself. In fact,
 by looking for the maximum slope y, we obtain simultaneously
 $f(x)$. The relationship between convexity and duality appears in many
 places in chemical physics. An example of such a relationship are the
 Legendre transforms in thermodynamics.}
\label{thermo}
\end{figure}

Duality appears in many places in chemical physics, albeit without this name. It is intimately connected with convexity~\cite{strang}. Multidimensional convex functions obey a range of useful properties,  one of which is that at each point there  is at least one hyperplane  tangent to the graph, whose slope is the derivative of that function.  In fact, this plane belongs to a family of planes  that separate the entire configurational space in two parts, below and above the function. Convexity ensures that they never cross the function in more than one point.   
 To demonstrate how duality is related to this,  we restrict ourselves to a 1 dimensional parabola 
$f(x)=x^2$ but the argument is completely general. The separating  planes now become straight lines (Fig.~\ref{thermo}). 
There is a set of lines with slope y which lie below the parabola as long as $yx-d\leq x^2$ where $d$ is the depth. The line will touch tangentially  the parabola when $d=1/4y^2$. Each point on the parabola can be represented as  a line which touches the parabola with different slopes y. 
 The envelope of these lines  reproduces the parabola itself. In fact, by looking for the maximum slope y,   we obtain simultaneously  $f(x)$.  This is in fact what the Legendre transforms do in thermodynamics  where  for example the  Gibbs and Helmholtz free energy are coupled with pressure and volume respectively. This interpretation was already  recognised by Tisza\cite{tisza}. The dual  emphasises the  geometric interpretation of the problem, and in that sense its variables  can be interpreted as generalised Lagrange multipliers. It further clarifies how we can obtain information about a function by looking {\it outside} and bounding it externally from below.

\section{Results} \label{Results}
To demonstrate the SOS  approach in physical applications, we apply it  now to   model  potentials that highlight the versatility of the method in dealing with typical aspects of landscapes that make them difficult to study. Once the problem is formulated theoretically as described above,  we use 
 third party $\textrm{Matlab}$ toolboxes to carry out the SOS decomposition: SOSTOOLS, a tool  for producing the SOS decompositions of polynomial optimisation problems \cite{prajna02} and 
SeDuMi~\cite{sturm01} or SDPT3~\cite{tutuncu01}, SDP solvers 
 which use  primal-dual interior point methods. Other minimisations for comparisons 
were implemented in the $\textrm{Matlab}$ optimization toolbox.

\subsection{Multi-well potential} 
 To demonstrate the method we apply it first to a potential with  many minima whose
values are very close to the global minimum but are separated by high barriers 
(see Fig. \ref{1D_multiwell}). The typical ratio of barrier height/local minima energy difference is 80. 
 In addition, the minimum closest in energy to the global minimum is furthest apart in 
domain space.  This type
of poor connectivity  behaviour  is not uncommon in potential energy surface
problems where a number of nodes on the disconnectivity graph\cite{becker97} have
energy levels that are almost identical.  

The SOS approach provides  the global minimum which in this case is proven to be exact from the duality gap,  and,  correctly identifies its location in the coordinate space (Table 1). 
Additionally, the SOS decomposition we obtained is a faithful representation of the
original function (Fig. \ref{1D_multiwell}). 
A general feature of the SOS optimisation, once formulated theoretically,  is that it is carried out once, unlike other methods which may  depend on a good initial guess or on the location of the initial conditions, and, it  provides all the information simultaneously.

We have compared  to two other common minimisation approaches:
the downhill simplex method\cite{nelder65} and the
BFGS quasi-newton scheme \cite{dennis83}, which is used in the basin-hopping algorithm. Unlike SOS,  both methods were applied iteratively from a grid of starting points spanning the domain space of the problem with the grid density being increased until the global minimum was obtained. In order to succeed in finding the global minimum, at least 10 separate runs
from starting points on an evenly spaced grid of points spanning the
domain were required. This is in fact more attempts than the
number of minima in the domain. The  point however remains that these algorithms  have no way of establishing that the
minimum that they locate  is in fact the global minimum.

We note that there is nothing special about the form of the potential presented here. We have tried successfully several multiwell examples with different polynomial representations, more asymmetric 
wells, and, separations with identical success.   
Interestingly, the method can identify the global minimum even if the wells are nearly-degenerate with arbitrarily close energy.  In fact, a  true degeneracy  is  reflected in the eigenvalues of the dual solution matrix  and  can hence be identified and located in the dual solution matrix. This can be very useful in locating often missed degenerate 
minima by looking at  the dual. 

The formalism holds for multi-dimensional problems and so a
generalisation of this problem to higher dimensions  works in the same
fashion. The global minimum ($U=-14.036$, location $=[-0.9798,-0.9798,-0.9798,-0.9798]$) is identified among on the order of $4100$ densely spaced minima (Fig.~\ref{4D_multiwell}) of the 4 dimensional version of the same potential. At the same time, a ten dimensional 
version  with  now disconnected minima dispersed in a higher 10
dimensional space (of the order of 59,000 minima) worked equally well (Table 1). 

\subsection{Golf-like  potential} 
A different set of energy landscapes  exhibit  flat regions with  deep and  narrow wells which become especially challenging and rapidly intractable for searching algorithms. The reason is that the flat regions provide no energy differences that can guide any search. Exhaustive searching is  the only option, which becomes intractable very rapidly as the domain increases.  We were able to successfully locate the global minimum of  a delta function embedded in a two-dimensional  infinite  region. 
 Such an example is shown in Fig. \ref{golf} 
 where the function is almost flat everywhere except in two locations where
 almost identical delta function wells are present. Note that we 
have included the entire $\mathbf{R}^2$ domain.  The SOS approach correctly identifies  the minimum and its location (Table 1). This is also true when  the higher energy  well becomes wider. 
 The  power of this approach 
comes from the following general simultaneous reasons: (i) the solution is always  approached  from below so 
the flatness of the function does not enter (ii)the minimum is very
 narrow in configurational space, however, in the lifted space of the
 coefficients this is no longer necessarily true  (iii) the set of equations linking the coefficients of the function intersect the cone of positive definite matrices. The possible solutions are now on the boundary of this space and no longer on the interior and hence easy to distinguish near degeneracies even if disconnected in configuration space.

\subsection{M\"uller potential}
 The M\"uller potential \cite{Mueller80} is commonly used as a non-trivial test for  reaction path methods  due to the complexity of 
its minima structure that leads to a highly contorted reaction coordinate which requires a sharp change of direction at the saddle.  The potential surface (Fig. \ref{mueller_fig}) contains three 
minima, with the global minimum of $-146.7$ at $(-0.558,1.442)$: 
\begin{eqnarray}U(x,y)=\sum_{i=1,\ldots,4}A_i\exp[a_i(x-x_i)^2+b_i(x
-x_i)(y-y_i)+c_i(y-y_i)^2]
\end{eqnarray}
where  $A=(-200, -100, -170, 15)$, 
 $a=(-1, -1, -6.5,0.7)$, $b=(0,0,11,0.6)$, $c = (-10, -10, -6.5, 
0.7)$, $x_i = (1,0,-0.5,-1)$, $y_i = (0,0.5,1.5,1)$.   Its domain is constrained in order to exclude areas where the function is not well behaved. 
 The SOS approach allows for constraints, both 
equality and inequality, to be incorporated on an equal footing 
as can be seen in Eq. 3 of  Fig. 1. We  can take  advantage of this powerful tool here by  excluding parts of space that are not of interest  in the formulation of the problem. 
The definition of this function includes an exponential so in order to
 apply the method here we fitted the function in a least squares sense
 to a polynomial of degree 16, $U_{f}$, which  contains 153
 monomials, 67 of which are odd. (The RMS error normalised by the data
 range was 0.0015 over the points sampled).
 This leads to  the minimisation of $U_f$ subject to a series of inequality constraints, $g_i$:
\begin{eqnarray}
\begin{array}{rl}
\textrm{max:} & \lambda \\
\textrm{ subject to:} & U_{f}-\sum_{i=1}^4 s_ig_i -\lambda \succeq 0 \\
 & g1: x\geq -1.5 \quad g2: x\leq 0.7 \\
 & g3: y\geq -0.5 \quad g4: y\leq 2.0, \\ \nonumber
\end{array}
\end{eqnarray}
where $s_i$ are additional SOS functions with unknown coefficients.   
 
The SOS approach succeeded in locating the minimum which is proven to be exact by the duality gap. It is interesting to point out that the final  SOS decomposition of the potential 
 is a good representation of the original  function (see Fig. \ref{mueller_fig}) which may allow for additional properties 
of the function to be explored in  different representations. For example, we were able to correctly locate the elusive saddle through eigenvector following on the SOS function. If more faithful representations were sought,  a better bound to the function can be produced by systematically  augmenting the dimension of the basis  (eq. 3, Fig 1) at higher computational cost.   

\section{Discussion} \label{Discussion}
The SOS approach\cite{parrilo_thesis00} is  a 
rigorous and systematic way to approaching complex  problems  by
exploiting both their algebraic and geometric properties.  We have
applied it here  to the identification of the  global minimum of model
potential energy surfaces which exemplify features that make this
problem difficult in physical systems of interest such as narrow and
disconnected wells embedded in flat or contorted regions. 
 We have demonstrated the method  on 
 low dimensional problems in which it is possible to gain an understanding of how the method
 works and how it is different  to other approaches. It is clear that no method will outperform all other methods for all  problems and it  may well be that heuristic approaches may
outperform the  more formal SOS approach  on a given
subset of problems. However, the appeal of the SOS approach stems from its generality  due to its deep theoretical foundations and efficient computational algorithms.

Another advantage comes from  the added physical insight that can be gained by formulating the problem in this way where duality holds. An immediate consequence of duality is  the ability to produce a proof when the exact answer has been reached without sampling or comparing to other minima, in contrast to almost all other methods. Secondly, the geometrical interpretation attached 
to the dual can lead to insight into the problem, including the identification of degeneracies and symmetries. 
Furthermore, there is freedom in how to pose a problem, so that constraints (equalities and inequalities) can be incorporated as variables or new variables can be introduced which can lead  
to  efficient answers.  This is particularly useful for dealing with more complex potentials that include trigonometric functions and so  fitting could be avoided by introducing  new variables. This may extend the applicability of the method to certain non-polynomial functions.

Unlike other methodologies that approximate the functions either by subtracting or adding extra terms,which is known to often
alter the behaviour of the system especially in multidimensional
problems, in this approach  the original function is never deformed nor linearised. Rather, an entirely different and higher dimensional
problem is sought, which is  both convex solvable and whose  solution
in the space of the original variables coincides with  that of the original
question. Unlike gradient-based and Monte-Carlo methods that use the original function, this approach bounds the function {\it externally} from below. This is one of the reasons that poorly connected and/or nearly degenerate minima separated by high 
barriers do  not pose the difficulties encountered in the usual sense.

As the complexity of problems attempted increases, it is possible that the exact solution may not be found. In many cases, a ``close enough'' answer may be adequate. Here, we can quantify the gap between our answer and the exact solution.  If the  lower bound solution is not adequately close, then further
refinement can be sought by seeking a higher dimensional monomial
basis or augmenting the equation of the Positivstellensatz by introducing SOS functions, which can be thought here  as a generalisation of Lagrange multipliers.  Accuracy
can be systematically increased at the expense of more computational
cost, since there is a proof of the existence of the exact answer in
such liftings to higher dimensions.  In that respect, 
this method is closer in spirit to a variational approach but one which is guaranteed to be optimal 
and may lead eventually, at least in principle, to the right answer of the original problem.  Unlike other global approaches, the method allows for  insight into the nature of the PES rather than just providing  a number. For example, having an explicit SOS polynomial representation of the problem can potentially be useful for studying other properties of the system besides its global minimum. 
  
A number of challenges remain due to the  significant scaling of the number of the new  variables that occurs as a result of the lifting process which  means that, for physical problems, 
 the current limitations on size of SDPs solvable using primal-dual 
interior point methods  need to be overcome. Excluding symmetries and symmetry operations is a promising way in that direction. This is the focus of ongoing research.

We  thank Pablo Parrilo \& Mauricio Barahona for
helpful discussions. This work has been funded by  the BBSRC, UK and
the ONR, USA.

\newpage

\bibliography{preprint}

\newpage

{\bf List of figures:} \\
\begin{enumerate}
\item
Convex duality in Legendre transforms.
\item
Schematic flowchart of a SOS optimisation.
\item
SOS optimisation on a 1D multiwell potential.
\item
A 2D projection of a 4 dimensional potential of $16^{th}$
degree.
\item
Contour plot of a Golf-like landscape.
\item
Contour plot of the original fitted M\"uller potential (a), and its SOS 
representation (b).
\end{enumerate}
\newpage

{\bf List of tables:}\\
\begin{enumerate}
\item
Summary of SOS optimisation results for the 1D, 10D multiwell,
M\"uller and Golf-like potentials.
\end{enumerate}
\newpage

{\bf List of figure captions:}
\\
Fig. 1: How convexity and duality are related. A set of lines with
slope y lie below the parabola $f(x)=x^2$, as long as $yx-d\leq x^2$ where $d$ is the depth. The line will touch tangentially  the parabola when $d=1/4y^2$. Each point on the parabola can be represented as  a line which touches the parabola with different slopes y.  The envelope of these lines  reproduces the parabola itself. In fact,
 by looking for the maximum slope y, we obtain simultaneously
 $f(x)$. The relationship between convexity and duality appears in many
 places in chemical physics. An example of such a relationship are the
 Legendre transforms in thermodynamics.
\\
Fig. 2: Schematic flowchart of a SOS  optimisation. The original
  global energy minimisation problem is reformulated in a series of
  steps: (1) a polynomial optimisation including any constraints is
  transformed to (2) a maximisation of a  real number $\lambda$  which will keep the function nonnegative. A sufficient but not necessary condition for this to be true is that the function is a SOS, so the problem is rewritten as (3). The constraints can be incorporated as well and algebraic geometry guarantees the validity of this equation. The additional terms provide a series of hierarchical approximations to the true solution. (4) A function being a SOS is equivalent to finding a positive semidefinite matrix Q where z is a vector of monomials of U. Note that Q relates the coefficients of U and contains  $\lambda$.  Finding Q is equivalent to the  semidefinite programme given in (5), with the  primal (left) and dual (right) problems. Each step is an exact reformulation of the previous  apart from (3) which is  guaranteed to give an optimal bound to the original problem.  When expressed in the appropriate monomial basis the problem of 
locating the coefficients of Q becomes convex and hence efficiently solvable. 
However, the SOS function (3) is still a non-convex function in configuration space and can be a good representation of the original function.
\\
Fig. 3: SOS  optimisation on a multiwell
  potential U(-) with 8 minima, the leftmost being
  the global minimum.  The second minimum,  farthest away in domain space, is very close in value
  to the global minimum. The
  global minimum is correctly identified. The SOS function $f^{SOS} (*)$ is 
in excellent agreement with U.   Here, 
$U=0.01x-128.1x^2+2688x^4+-21504x^6+84480x^8 -180224x^{10}+212992x^{12}-131072x^{14}+32768x^{16}$.
\\
Fig. 4: A 2D projection of a 4 dimensional multiwell
  potential with 4096 minima given by:
  $U(x_1,x_2,x_3,x_4)=\sum_{i=1}^4a_0+a_1x_i+a_2x_i^2+a_3x_i^4+a_4x_i^6+a_5x_i^8+a_6x_i^{10}+a_7x_i^{12}+a_8x_i^{14}+a_9x_i^{16}$
  where $a_0=1, a_1=0.5, a_2=-130.1, a_3=2688, a_4=-21504, a_5=84480, a_6=-180224, a_7=212992, a_8=-131072, a_9=32768$. SOS optimization sucessfully identified
  the global minimum of $U=-13.977$ to within 0.026 at the point $[-0.9792,-0.9792,-0.9792,-0.9792]$.
\\
Fig. 5: Contour plot of a
  Golf-like potential. U is given by $\frac{-1}{\pi}\left(\frac{e_1}{(x-5)^2+(y-4)^2+e_1^2}+\frac{e_2}{(x+7)^2+(y+6)^2+e_2^2}\right),e_1=0.09, e_2=0.1$
  defined over the entire $\mathbf{R}^2$ domain (left). The area close
  to the origin (blue circle) is reproduced at greater
  magnification on the right. The two minima are $U(5,4)= -3.5369$ (global) 
and   $ U(-7,6)= -3.1832$. In this case the SOS optimisation produces a
  lower bound of $U= -3.539$. The non-negligible duality gap (Table 1)  
shows that the exact answer is at most $0.0023$ away. The configurational coordinates are obtained correctly to  within numerical accuracy.
\\
Fig. 6. Contour plot of the original fitted M\"uller potential (a), and its SOS 
representation (b). The global minimum was identified correctly.  $f^{SOS}$  (b) is the projection to the original domain space from the
lifted space of the SOS function  including the inequality constraints. Much of the original function is preserved within the first order of  Eq. 3, Fig. 1.
\newpage


\begin{table}
\begin{center}
\begin{tabular}{|l|c|c|c|c|}
\hline
                & {\bf 1D Multi-well} & {\bf 10D Multi-well}  & {\bf M\"uller} & {\bf Golf-like}\\
\hline
{\bf CPU time (s)}	& 0.65     & 6840.00       & 10.06               & 0.80\\
{\bf Lower bound}	& -2.106 [-2.106] & -193.685 & -146.674 [-146.674] & -3.539 [-3.537] \\
{\bf Configuration}	& -0.980          & see below (*)  & (-0.557,1.449)      & (4.994,3.995)  \\
{\bf Duality gap}	& 3.857E-06       & 7.56E-04   & 1.366E-4            & 0.002   \\
{\bf Primal}		& 2.105998        & 193.6848 & 146.6743           & 3.537  \\
{\bf Dual}		& 2.106002  	  & 193.6855 & 146.6745           & 3.539  \\
\hline
\end{tabular}
\end{center}

\caption[Summary of SOS optimisation results for 
the  1D, 10D multiwell, M\"uller and Golf-like
potentials.]{Summary of SOS optimisation results for 
the  multi-well, M\"uller and golf-like potentials: the 
 cpu time in a standard PC in seconds, the lower
bound value with the  exact global minimum in brackets, the configurational 
coordinates of the minimum, the duality gap, and the  final values of the primal and 
dual problem. The 10 dimensional multiwell example was defined as follows:
$ U  =\sum_{i=1}^{10}
a_1x_i+a_2x_i^2+a_3x_i^3+a_4x_i^4+a_5x_i^5+a_6x_i^6 $
where $a_1=2, a_2=230, a_3=-28, a_4=-1000, a_5=1, a_6=1000 $ and the global
minimum of $-193.685$ was located among approximately 59000 minima at $(*)[0.7369, 0.7369, 0.7369, 0.7369,
  0.7369, 0.7369, 0.7369, 0.7369, 0.7369, 0.7369]$.}

\label{results_summary}
\end{table}


\begin{figure}[h!]
\includegraphics[width=8.5cm]{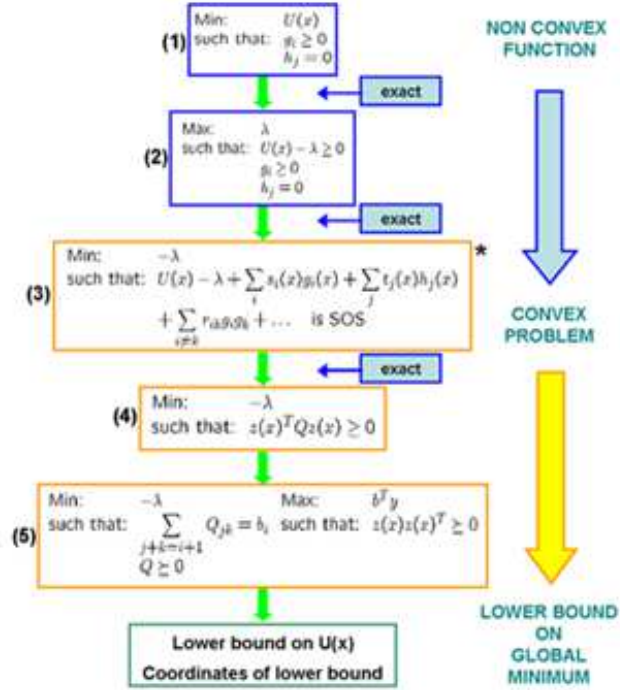}
\caption[Schematic flowchart of a SOS  optimisation.]{Schematic flowchart of a SOS  optimisation. The original
  global energy minimisation problem is reformulated in a series of
  steps: (1) a polynomial optimisation including any constraints is
  transformed to (2) a maximisation of a  real number $\lambda$  which will keep the function nonnegative. A sufficient but not necessary condition for this to be true is that the function is a SOS, so the problem is rewritten as (3). The constraints can be incorporated as well and algebraic geometry guarantees the validity of this equation. The additional terms provide a series of hierarchical approximations to the true solution. (4) A function being a SOS is equivalent to finding a positive semidefinite matrix Q where z is a vector of monomials of U. Note that Q relates the coefficients of U and contains  $\lambda$.  Finding Q is equivalent to the  semidefinite programme given in (5), with the  primal (left) and dual (right) problems. Each step is an exact reformulation of the previous  apart from (3) which is  guaranteed to give an optimal bound to the original problem.  When expressed in the appropriate monomial basis the problem of 
locating the coefficients of Q becomes convex and hence efficiently solvable. 
However, the SOS function (3) is still a non-convex function in configuration space and can be a good representation of the original function.}
\label{flowchart}
\end{figure}

\begin{figure}[h!]
\includegraphics[width=8.5cm]{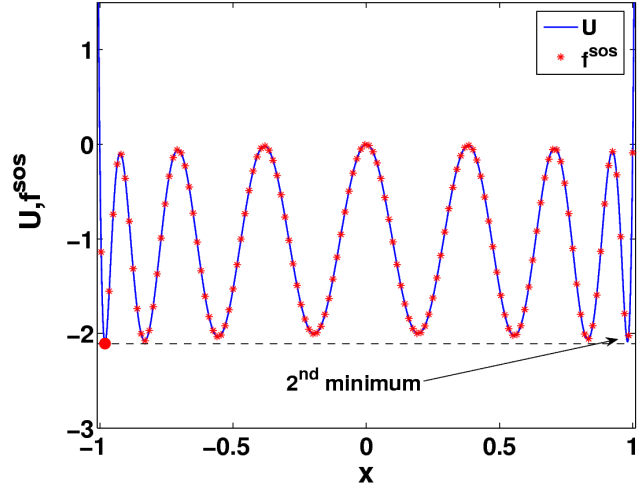}
\caption[SOS  optimisation on a  1D multiwell
  potential.]{SOS  optimisation on a multiwell
  potential U(-) with 8 minima, the leftmost being
  the global minimum.  The second minimum,  farthest away in domain space, is very close in value
  to the global minimum. The
  global minimum is correctly identified. The SOS function $f^{SOS} (*)$ is 
in excellent agreement with U.   Here, 
$U=0.01x-128.1x^2+2688x^4+-21504x^6+84480x^8 -180224x^{10}+212992x^{12}-131072x^{14}+32768x^{16}.$}
\label{1D_multiwell}
\end{figure}


\begin{figure}[htbp!]
\includegraphics[width=8.5cm]{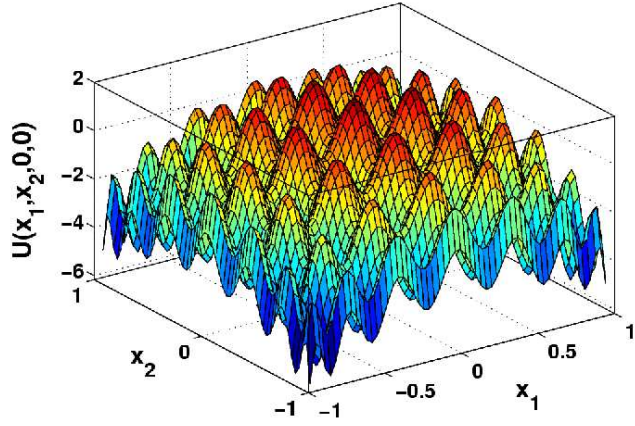}
\caption[A 2D projection of a 4 dimensional potential of 16^{th}
  degree.]{A 2D projection of a 4 dimensional multiwell
  potential with 4096 minima given by:
  $U(x_1,x_2,x_3,x_4)=\sum_{i=1}^4a_0+a_1x_i+a_2x_i^2+a_3x_i^4+a_4x_i^6+a_5x_i^8+a_6x_i^{10}+a_7x_i^{12}+a_8x_i^{14}+a_9x_i^{16}$
  where $a_0=1, a_1=0.5, a_2=-130.1, a_3=2688, a_4=-21504, a_5=84480, a_6=-180224, a_7=212992, a_8=-131072, a_9=32768$. SOS optimization sucessfully identified
  the global minimum of $U=-13.977$ to within 0.026 at the point $[-0.9792,-0.9792,-0.9792,-0.9792]$.}
\label{4D_multiwell}
\end{figure}




\begin{figure}[htbp!] 
\begin{center}
\includegraphics[width=8.5cm]{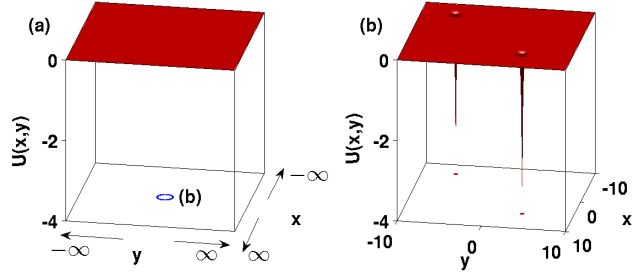}
\end{center}
\caption[Contour plot of a Golf-like landscape.]{Contour plot of a
  Golf-like potential. U is given by $\frac{-1}{\pi}\left(\frac{e_1}{(x-5)^2+(y-4)^2+e_1^2}+\frac{e_2}{(x+7)^2+(y+6)^2+e_2^2}\right),e_1=0.09, e_2=0.1$
  defined over the entire $\mathbf{R}^2$ domain (left). The area close
  to the origin (blue circle) is reproduced at greater
  magnification on the right. The two minima are $U(5,4)= -3.5369$ (global) 
and   $ U(-7,6)= -3.1832$. In this case the SOS optimisation produces a
  lower bound of $U= -3.539$. The non-negligible duality gap (Table 1)  
shows that the exact answer is at most $0.0023$ away. The configurational coordinates are obtained correctly to  within numerical accuracy.} 
\label{golf}
\end{figure}


\begin{figure}[h!]
\includegraphics[width=8.5cm]{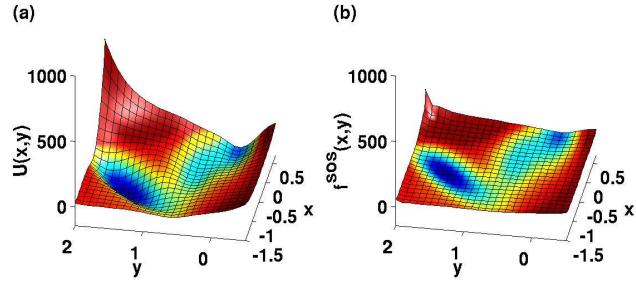}
\caption[Contour plot of the original fitted M\"uller potential (a), and its SOS 
representation (b).]{Contour plot of the original fitted M\"uller potential (a), and its SOS 
representation (b). The global minimum was identified correctly.  $f^{SOS}$  (b) is the projection to the original domain space from the
lifted space of the SOS function  including the inequality constraints. Much of the original function is preserved within the first order of  Eq. 3, Fig. 1.}
\label{mueller_fig}
\end{figure} 

\end{document}